\documentclass{PoS}

\title{Current status of LEGEND: Searching for Neutrinoless Double-Beta Decay in $^{76}$Ge: Part II}

\ShortTitle{LEGEND Status}

\author{J.M.L\'opez-Casta\~no \thanks{Speaker of ``Current Status of LEGEND''.}\\
        University of South Dakota\\
        E-mail: \email{Mariano.Lopez@usd.edu}}
\author{I.Guinn \thanks{Speaker of ``LEGEND: Searching for Neutrinoless Double-Beta Decay in $^{76}$Ge''.}\\
        University of North Carolina\\
        E-mail: \email{iguinn@uw.edu}}

\abstract{Neutrinoless double-beta decay($0\nu\beta\beta$) decay is a hypothetical process that violates lepton number, and whose observation would unambiguously indicate that neutrinos are Majorana fermions. In the standard inverted-ordering neutrino mass scenario, the minimum possible value of m$_{\beta\beta}$ corresponds to a half-life around 10$^{28}$ yr for $0\nu\beta\beta$ decay in $^{76}$Ge, which is the target of the next generation of experiments. The current limits of GERDA and \textsc{Majorana Demonstrator} indicate a half-life higher than 10$^{26}$ yr. These experiments use high-purity germanium (HPGe) detectors that are highly enriched in $^{76}$Ge. They have achieved the best intrinsic energy resolution and the lowest background rate in the signal search region among all $0\nu\beta\beta$ experiments.\\\\
Taking advantage of these successes, a new international collaboration - the Large Enriched Germanium Experiment for Neutrinoless $\beta\beta$ Decay (LEGEND) - has been formed to build, following a phased approach, a ton-scale experiment with discovery potential covering the inverse-ordering neutrino mass range in a decade. 
The first part of LEGEND proceedings describes GERDA and \textsc{Majorana Demonstrator} capabilities and the general plan of LEGEND to reach the goal, while this second part is focused in the status of the first stage of LEGEND, LEGEND-200.
}

\FullConference{XXIX International Symposium on Lepton Photon Interactions at High Energies - LeptonPhoton2019\\
		August 5-10, 2019\\
		Toronto, Canada}

\begin{document}

\begin{figure}[t]
\begin{center}
  \includegraphics[height=4.5cm]{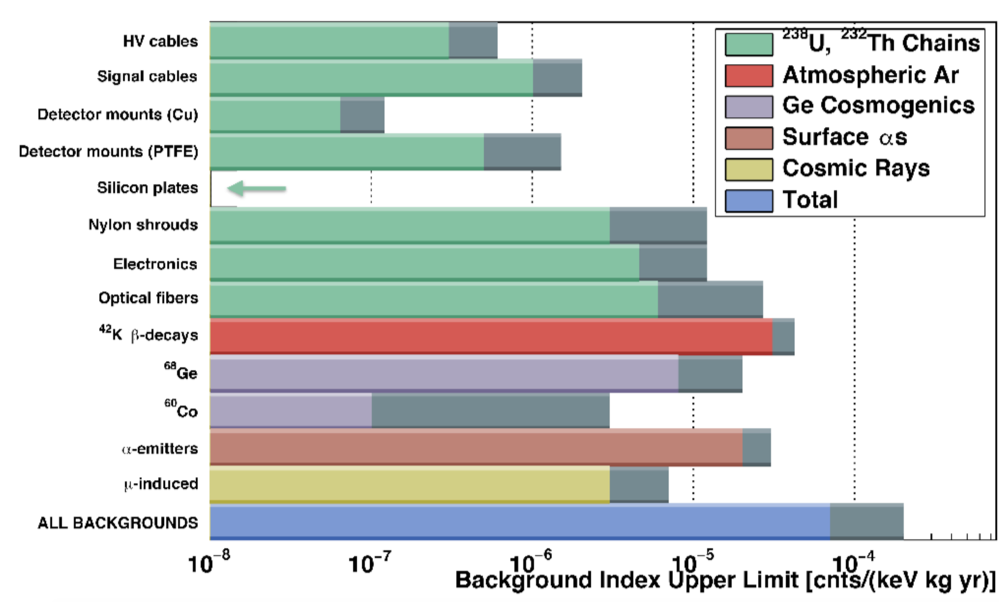}
  \includegraphics[height=4.5cm]{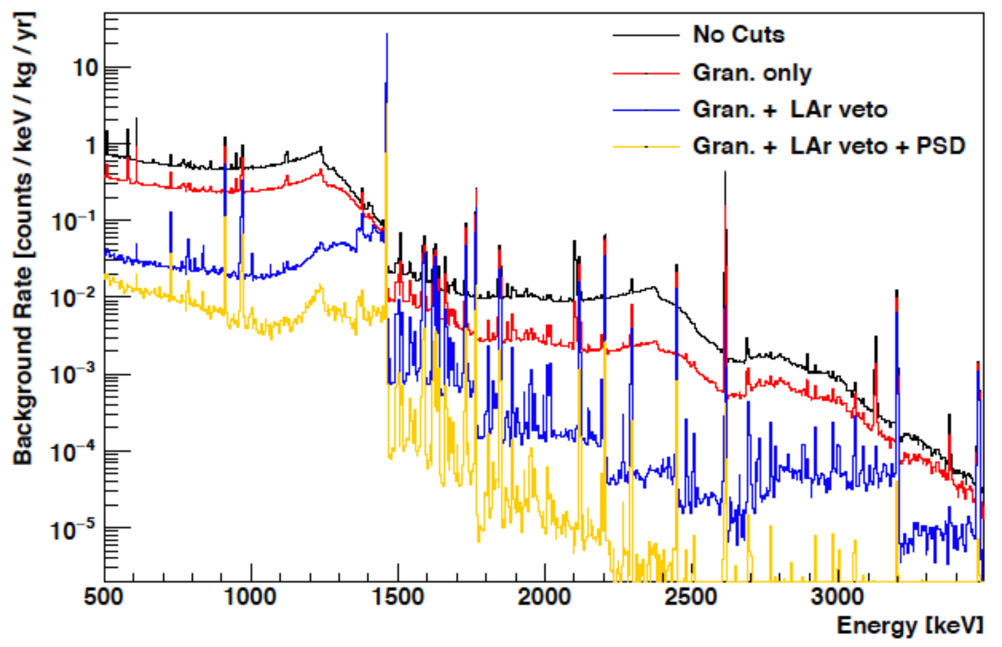}
  \caption{ 
 Monte Carlo simulations based on experimental data and material assays. Left: Background rates at 90\% CL. The color bar indicates the type of background present in the element. The grey bar indicates the uncertainty in overall background rejection efficiency. Right: Energy spectrum before cuts (black), after granularity cut (red) after granularity and LAr veto (blue) and with all cuts applied (yellow) \cite{Steve}.}
  \label{Back}
  \end{center}
\end{figure}

\section{The LEGEND experiment}

Combining the best technologies of GERDA \cite{GERDA}\cite{GERDA2} and \textsc{Majorana Demonstrator} \cite{MJD}\cite{MJD2}, and including contributions of other groups that are not present in the previous experiments, the Large Enriched Germanium Experiment for Neutrinoless $\beta\beta$ Decay (LEGEND) \cite{LEGEND} aims to reach a sensitivity of 10$^{28}$~yr for neutrinoless double beta ($0\nu\beta\beta$) decay half-life ($T^{0\nu}_{1/2}$) in $^{76}$Ge.
The requirement to reach that goal is at least 10~t-y of isotopic exposure in High Purity Germanium (HPGe) detectors, and a background level lower than 0.1 cts/(FWHM t y).

LEGEND will have 2 different phases; in the first phase (LEGEND-200), 200 kg of HPGe detectors will be deployed in the existing GERDA infrastructure with a background goal of $<0.6$~cts/(FWHM t yr).
This will provide a half-life sensitivity of $\sim10^{27}$~y after 5 years of data taking.
LEGEND-200 is in its final stage of design and beginning construction stage, with data taking projected to begin in 2021.
The second phase, LEGEND-1000, will operate 1000 kg of HPGe detectors.
This last stage is in the early planning and R\&D phase.

\section{LEGEND-200}

LEGEND-200 will implement several improvements for achieving further background reductions and assuring a good detector operation.
For example, additional cables and electronics and a new vacuum feedthrough will be introduced to accommodate additional detectors in the GERDA infrastructure.
Also, a mini-shroud covering the HPGe detector surfaces has been introduced and tested successfully to reduce the number of background events coming from the lithium-doped surface. Those events are produced by $\beta$ decay of $^{42}$K deposited in the crystal surface by the LAr \cite{K42}.

\begin{figure}[t]
\begin{center}
  \includegraphics[height=3cm]{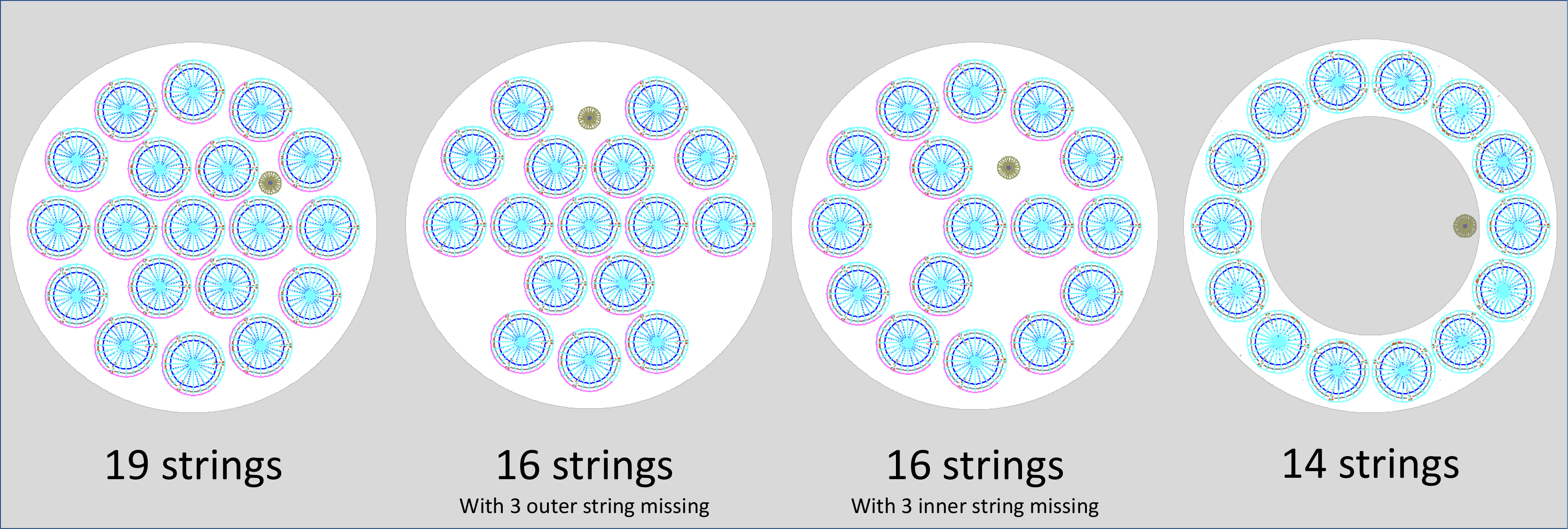}
  \caption{ 
  Top view of the four main array geometries considered. The blue circles represent detector strings and the small green circle the possible location of 1 calibration source (if more than 1 source is used, the other locations will be determined by symmetry) \cite{array}.}
  \label{Array}
  \end{center}
\end{figure}

LEGEND-200 will reuse 20~kg of enriched Broad Energy Germanium detectors, 15~kg of enriched semi-coaxial detectors and 10~kg of Inverted Coaxial Point-Contact detectors (ICPC) from GERDA, and 28~kg of enriched P-type Point-Contact detectors from the \textsc{Majorana Demonstrator}.
In addition, around 150 kg of new ICPC enriched detectors are being manufactured and characterized for use in LEGEND-200\cite{Detectors}. 
The expected background has been calculated from the experimental data and material assays, showing an expected reduction of a factor 5 in relation to GERDA and \textsc{Majorana Demonstrator}, as Figure \ref{Back} shows.
Several geometries have been considered for the array geometry, as Figure \ref{Array} shows.
Simulations also show that geometries with a high number of arrays inside the ring have a degradation in the liquid argon (LAr) rejection capabilities.
For that reason, currently, the ring of 14 strings and variations with only a few string inside are the preferred by the collaboration \cite{array}.
The string arrangement further impacts the calibration system design, which must position a calibration source within the array to illuminate all detectors.

\section{Conclusions}
The expected background index for LEGEND-200 is estimated by simulations from the measured materials radio-purity.
These simulations show a reduction in backgrounds by a factor 5 \cite{Steve} in relation to GERDA and the \textsc{Majorana Demonstrator}.
Based on simulations and the results of GERDA and the \textsc{Majorana Demonstrator}, LEGEND-200 will meet its background target of $<0.6$~cts/FWHM-t-y.
Expected improvements from active R\&D on LEGEND-1000 indicate that the accuracy on the energy determination and the effectiveness of background suppression techniques will reduce the background index to the levels required to achieve a sensitivity to the $0\nu\beta\beta$ half-life of 10$^{28}$ yr at 90\% CL.

\end{document}